\documentclass{article}

\title {Yet Another Efficient Unification Algorithm}
\author {Alin Suciu }
\date {}
\begin{document}
\maketitle
\begin{center}
 Department of Computer Science

 Technical University of Cluj-Napoca

 26-28, George Baritiu St., RO-3400, Cluj-Napoca, Romania

 Tel/Fax: +40-64-194491

 Alin.Suciu@cs.utcluj.ro
\end{center}

\abstract { The unification algorithm is at the core of the logic
programming paradigm, the first unification algorithm being
developed by Robinson [5]. More efficient algorithms were
developed later [4] and I introduce here yet another efficient
unification algorithm centered on a specific data structure,
called the Unification Table.}

\section{Introduction}

The unification algorithm is at the heart of the logic programming
paradigm [3]. Starting with the classic algorithm of Robinson [5],
the unification algorithm was developed to become more and more
efficient [4]. Even nowadays the unification theory is still under
development and is receiving continuous scrutiny from the
scientific community [2].

The present paper presents yet another efficient unification
algorithm centered on a data structure called Unification Table,
which borrows some ideas from the data structures used by the
Warren's Abstract Machine [1].

The next paragraph presents in detail the proposed unification
algorithm, giving the C-style pseudo code. An example of
application of the algorithm taken from [1] is also presented.

\section{The Unification Algorithm}

The unification algorithm below will compute the MGU of two
logical terms $x$ and $y$ if the unification is possible,
otherwise will report failure. As in all Prolog implementations of
the unification, the "occur check" is omitted; however, with some
modifications, the algorithm can implement a "unify with occur
check" procedure.

The algorithm is centered on a data structure called the Unification Table
(UT) which contains information related to each subterm (constant, variable,
composite) that occurs in the terms to be unified. The properties of the
unification table are crucial to the correctness and the efficiency of the
unification and will be described in detail below.

The algorithm consists of two main steps:

Step 1. Parse terms $x$ and $y$ and build the Unification Table (UT)

Step 2. Call the unification function \textit{unify(index(x), index(y))} where \textit{index(x)} and \textit{index(y)} are the indexes
of $x$ and $y$ in the UT.

In the following we will take a closer look to each step of the
unification algorithm, providing detailed explanation wherever
necessary.

\subsection{Step 1}

The idea of this step is to build a data structure, called
Unification Table (UT), in which every variable appears only once,
and all the subterms of p and q are included. The UT contains
three types of entries: variables (type VAR, arity 0), constants
(type STR, arity 0) and composite terms (type STR, arity greater
than 0).

The structure of the UT for each type of entry is the following:

\begin{table}[htbp]
\begin{tabular}
{|p{62pt}|p{42pt}|p{92pt}|p{81pt}|p{43pt}|p{87pt}|}
\hline
Term&
Index&
\textbf{Main functor}&
\textbf{Type of term (VAR/STR)}&
Arity&
\textbf{List of components} \\
\hline
variable&
i&
name of the variable&
VAR&
0&
Empty \\
\hline
constant&
j&
name of the constant&
STR&
0&
Empty \\
\hline
composite term&
k&
name of the main functor&
STR&
a&
list of the indexes of components \\
\hline
\end{tabular}
\label{tab1}
\end{table}

\noindent where:

Term - is the actual term; this column is never built; its purpose is only
to simplify the explanations and the understanding of the algorithm. We can
think of it as the result of a "write" procedure called upon the index of
the term.

Index - is the index of the table entry for some term; starts with 0 and
uniquely identifies the term.

Main functor -- is the main functor of the term.

Type is VAR for variables and STR for constants and composite terms.

Arity -- is the arity of the term; for variables and constants, it is 0.

For variables and constants, the list of components is the empty list.

For composite terms, the list of components is the sequence of the indexes
of the component subterms; the order is important -- it can be either left
to right or vice versa but not both of them. The parsing procedure is left
to right and ensures that when an entry is created for a composite term, all
its subterms are already in the UT. The parsing starts with the second term
($y)$.

Consider the following example from Ait-Kaci [1] requiring to find
the MGU of:

$x$ = p(Z,h(Z,W),f(W))

$y$ = p(f(X),h(Y,f(a)),Y)

The unification table will be:

\begin{table}[htbp]
\begin{tabular}
{|p{106pt}|p{49pt}|p{49pt}|p{86pt}|p{43pt}|p{77pt}|}
\hline
\textbf{Term}&
Index&
\textbf{Main functor}&
\textbf{Type of term (VAR/STR)}&
Arity&
\textbf{List of components} \\
\hline
Y&
0&
Y&
VAR&
0&
 \\
\hline
a&
1&
a&
STR&
0&
 \\
\hline
f(a)&
2&
f&
STR&
1&
1 \\
\hline
h(Y,f(a))&
3&
h&
STR&
2&
0 2 \\
\hline
X&
4&
X&
VAR&
0&
 \\
\hline
f(X)&
5&
f&
STR&
1&
4 \\
\hline
p(f(X),h(Y,f(a)),Y)&
6&
p&
STR&
3&
5 3 0 \\
\hline
W&
7&
W&
VAR&
0&
 \\
\hline
f(W)&
8&
f&
STR&
1&
7 \\
\hline
Z&
9&
Z&
VAR&
0&
 \\
\hline
h(Z,W)&
10&
h&
STR&
2&
9 7 \\
\hline
p(Z,h(Z,W),f(W))&
11&
p&
STR&
3&
9 10 8 \\
\hline
\end{tabular}
\label{tab2}
\end{table}

The table is filled by parsing the terms $x$ and $y$ from right to
left starting with $y$ in a bottom up manner. Each variable has
exactly one entry in the table but same constants or composite
terms may have different entries. The list of components for a
given term consists of the indexes of its components. The
unification function will start with the indexes of $x$ and $y$,
that is 6 and 11, so the call will be \textit{unify(6, 11)}.

\subsection{Step 2}

The \textit{unify} function called in Step 2 starts with the
indexes of $x$ and $y$ and uses two stacks \textit{Sx} and
\textit{Sy}. Initially the index of $x$ and $y$ are pushed on the
stacks \textit{Sx} and \textit{Sy} respectively. Then a main loop
will start that will continue until both stacks are empty. The
algorithm ensures that both stacks will be empty simultaneously
and that the stacks will eventually become empty if the
unification table was built correctly and the terms follow the
correct syntax for logic terms. The function will return either
SUCCESS or FAIL; in case of success each variable included in the
MGU will be marked using a global data structure. The pseudo code
for the \textit{unify} function follows:

\begin{verbatim}
function unify(ix, iy)
{
initialize stack Sx
initialize stack Sy
push ix on stack Sx
push iy on stack Sy
while (not_empty(Sx) and not_empty(Sy))
   {// start main loop
   pop i from Sx
   pop j from Sy
   // case 1: i is bound to a term and j is bound to a term}
   if (type(i) == STR and type(j) == STR)
      if (main functors of i and j match (both name and arity))
         if (arity > 0)
            push components of i on Sx in sequence
            push components of j on Sy in sequence
      else
         return(FAIL) // report failure
   // case 2: i is bound to a term and j is bound to a variable
   if (type(i) == STR and type(j) == VAR)
      if (j is a free variable)
         bind j to i and set mgu[j] = 1
      else // j is bound
         dereference j
         if (j is bound to a STR)
            push i on Sx
            push j on Sy
         else // j is bound to a free variable
            bind j to i
   // case 3: i is bound to a variable and j is bound to a term
   if (type(i) == VAR and type(j) == STR)
      // perfectly symmetric to case 2
   // case 4: i is bound to a variable and j is bound to a variable
   if (type(i) == VAR and type(j) == VAR)
      if (i is free and j is free)
         bind i to j (or vice versa) and set mgu[i] = 1
      if (i is free and j is bound)
         bind i to j and set mgu[i] = 1
      if (i is bound and j is free)
         bind j to i and set mgu[j] = 1
      if (i is bound and j is bound)
         push the index of the term to which i is bound on Sx
         push the index of the term to which j is bound on Sy
   } // end main loop
return(SUCCESS)
} // end function unify
\end{verbatim}

In order to help the understanding of the algorithm, C++ style
comments in italics are provided. The MGU for the above example
computed with this function is:

W=f(a)

X=f(a)

Y=f(f(a))

Z=f(f(a))

As one can see from the above function, the main loop extracts
elements i and j from the two stacks and then appropriate action
is taken according to the four possible cases.

In the first case i and j are indices of two elements of type STR
so the natural action to do here is to check if they have the same
name and arity. If so, in case the arity is greater than zero we
push all the components of i and j on the two stacks, otherwise we
report failure.

Cases two and three are symmetric and consider the case when
either i or j are variables but not both of them. In this case we
have to discriminate between the cases when the variable is free
(we just have bind it) or bound (in this case we have to
dereference the variable and push the result).

Finally in case four we have to deal with two variables and again
we have to consider four subcases according to the status of the
variables: free or bound. As before, if at least one variable is
free we just have to bind it, otherwise we must push the terms to
which the variables are bound on the stacks and continue the loop.

If the "while" loop ends naturally, without the forced return in
case of failure, then the function returns "success" as well as
the MGU.

In the case of an "occur check" violation the algorithm will
succeed, but an attempt to print the result will result in an
infinite loop, or a memory overflow; this can be solved as in many
Prolog systems by using a "guarded write" which will go only (say)
ten levels in depth.

\section{Conclusions}

The quest for efficient unification algorithms is the foundation
of increasing the efficiency of logic programs. Since the first
unification algorithm of Robinson [5], continuing with the
efficient algorithm of Martelli and Montanari [4], this quest
never stopped.

The unification algorithm presented in this paper is yet another
attempt to increase the efficiency of the unification. The
algorithms also benefits of simplicity and clarity which makes it
very easy to understand and implement.

An implementation in C is available as well as a Java applet.
Further developments of the algorithm are: a recursive version of
the algorithm and a version with "occur check".

\section{References}
\begin{description}
\item [[1]] Ait-Kaci, H., Warren's Abstract Machine, A Tutorial
Reconstruction, MIT Press, 1991.

\item [[2]] Baader, F., Snyder, W., Unification Theory, Handbook
of Automated Reasoning, A. Robinson, A. Voronkov eds., Elsevier
Science Publisher B.V., 1999.

\item [[3]] Lloyd, J.W., Foundations of Logic Programming,
Springer Verlag, 1984.

\item [[4]] Martelli, A., Montanari, U., An Efficient Unification
Algorithm, ACM Transactions on Programming Languages and Systems
(TOPLAS), Volume 4, Issue 2, April 1982, ACM Press New York, NY,
USA.

\item [[5]] Robinson, J.A., A Machine-oriented Logic Based on the
Resolution Principle, JACM, 12, 1, Jan 1965, 23-41.
\end{description}
\end{document}